\documentclass[fleqn,usenatbib]{mnras}

\usepackage{newtxtext,newtxmath}

\usepackage[T1]{fontenc}

\DeclareRobustCommand{\VAN}[3]{#2}
\let\VANthebibliography\thebibliography
\def\thebibliography{\DeclareRobustCommand{\VAN}[3]{##3}\VANthebibliography}

\usepackage{graphicx}	
\usepackage{amsmath}	

\title[AIC and CMIC inside PNe to form NSs]{Formation of neutron stars via accretion-induced collapse and core-merger-induced collapse inside planetary nebulae from white dwarf binaries}

\author[I. Ablimit]{
Iminhaji Ablimit,$^{1,2,3}$\thanks{I already left all these affiliations.}
\\
$^{1}$Department of Astronomy, Kyoto University, Kitashirakawa-Oiwake-cho, Sakyo-ku, Kyoto 606-8502, Japan\\
$^{2}$Henan Academy of Sciences, Zhengzhou, Henan 450046, China\\
$^{3}$CAS Key Laboratory for Optical Astronomy, National Astronomical Observatories, Chinese Academy of Sciences, Beijing 100012, China
}

\pubyear{2026}

\begin{document}
\label{firstpage}
\pagerange{\pageref{firstpage}--\pageref{lastpage}}
\maketitle

\begin{abstract}
The accretion-induced collapse (AIC) and  core-merger-induced collapse (CMIC) from oxygen-neon-magnesium (ONeMg) white dwarf (WD) binaries inside planetary nebulae (PNe) have not been previously even mentioned in the literature. In this paper, I propose and study two possible WD binary channels for AIC and CMIC to form neutron stars (NSs) within PNe. First, using simulations performed with the \textsc{MESA} stellar evolution code, I present a scenario in which NSs form via the AIC of ONeMg WDs inside PNe--referred to here as symbiotic nebulae. In the late evolutionary stages of ONeMg WD-red giant (or asymptotic giant branch) star binaries, substantial mass loss can occur through strong stellar winds or/and Roche-lobe overflow, potentially leading to the formation of nebulae surrounding central accreting WD binaries. These nebulae may be ionized by the hot cores of the giant stars or by the accreting WDs themselves. Under such conditions, the accreting WD may grow in mass to the Chandrasekhar limit and undergo collapse into a neutron star. NSs formed via this AIC channel are likely to retain WD companions, resulting in NS-WD binary systems, of which the Milky Way may host dozens. Second, through binary population synthesis modeling, I introduce another evolutionary pathway: the CMIC occurring during the common envelope evolution of ONeMg WD binaries. This process may result in the formation of a newborn NS within a PN -or, in some cases, a pulsar wind nebula.   
\end{abstract}

\begin{keywords}
(stars:) binaries (including multiple): close -- stars: evolution -- (stars:) white dwarfs -- (ISM:) planetary nebulae: general -- (stars:) supernovae: general - stars: late-type
\end{keywords}



\section{Introduction}

Planetary nebulae (PNe) and their central stars are very interesting objects that play crucial roles for improving our understanding of stellar and binary evolution.
PNe are believed to be products of the late evolutionary phase of zero age main sequence (ZAMS) stars with masses in the range of $\simeq 0.8 - 8.0 M_\odot$.
When these low- and intermediate-mass stars evolve to the phases of red giant (RG) or asymptotic giant branch (AGB),
a large amount of mass loss from the envelope of the RG or AGB stars may form nebulae, and hot (effective temperature of $\geq 3\times10^4$ K) and degenerate cores as the central objects can ionize the expanding nebulae to form PNe (e.g., \citealt{Kwok1983, TweedyKwitter1994, Soker2006, GuerreroDe Marco2013, Hillwigetal2017, KwitterHenry2022}).

Central stars of $\sim$20\% of known PNe are reported to be close binaries (e.g., \citealt{Bond2000, Miszalskietal2009, Ablimitetal2024}), the number of known cases is increasing with deeper sky surveys like Kepler and Gaia (e.g., \citealt{Bakeretal2018, BoffinJones2019, Jacobyetal2021, Chornay2021, ChornayWalton2022}), and see \cite{jones2017} for a Catalog at https://www.drdjones.net/bCSPN.
Mass-accreting WD binaries have been already observed as the central stars in some known PNe (\citealt{Bodeetal1987, Guerreroetal2004, Wessonetal2008, Munarietal2013, MaitraHaberl2022}). Some WDs might accrete mass from their non-degenerate companions (\citealt{Hamannetal2003, Guerreroetal2019}), a process that might explain  some puzzles, like the luminosity function of PNe (\citealt{Ciardullo2016, Davisetal2018, Souropanis2023}).


In this study, I propose two novel pathways for the formation of neutron stars (NSs) within planetary nebulae (PNe) through binary evolution. First, I investigate the evolution of white dwarf (WD)-giant star binaries using numerical simulations. Employing the \textsc{MESA} stellar evolution code, I model WD-red giant (RG) systems that undergo stable mass transfer and discuss how such evolution may lead to the formation of an NS within a PN in Section 2. Second, I explore an alternative formation channel involving mergers between WDs and the hot cores of giant stars within a common envelope. This scenario is examined systematically using binary population synthesis methods, as presented in Section 3. Finally, the conclusions of this study are summarized in Section 4.


\section{Formation of neutron stars via accretion-induced collapse of WDs inside symbiotic nebulae}
\label{sec:sec}

In the accretion-induced collapse (AIC) scenario, an oxygen-neon-magnesium (ONeMg) WD grows toward the Chandrasekhar mass by stably accreting hydrogen- and/or helium-rich material from a non-degenerate companion star. In addition to this AIC pathway, two WDs in a compact binary where the total mass exceeds the Chandrasekhar limit and at least one WD is of the ONeMg type may merge and collapse into a NS through what is known as merger-induced collapse (MIC). During the AIC/MIC, a deflagration involving oxygen and neon is expected to occur, but the nuclear energy released is insufficient to disrupt the tightly bound core (\citealt{miyaji1980}). As a result, ongoing electron captures eventually trigger gravitational collapse, leading to the formation of a NS (e.g., \citealt{nomoto1991}). \cite{bearsoker2021} mentioned formation of NSs after PNe when the WD inside its companion's envelope (obviously it is very different from the AIC discussed here), but they claimed it is impossible and the focus of their study is on the core collapse supernova explosion of He core which might occur inside a PN only if a third star forms the PN. Thus, AIC/MIC with PNe are not even considered seriously before.

The stellar wind mass loss or/and stable (or unstable) mass transfer in the evolution of oxygen-neon-magnesium (ONeMg) composition WD -- RG or ONeMg WD-AGB binaries have decisive roles on the formation of neutron stars (NSs) or/and PNe (symbiotic nebulae; see Figure 1 for the possible associated evolutionary pathways).
In a close WD binary with an RG or AGB companion star, there is a possibility that the giant companion could lose a significant amount of its mass through the stellar wind or/and during the mass accretion process. and this large amount
of mass loss may surround the binary system and form a nebula-like structure. If the nebula with the central accreting
WD system were formed, it could be ionized by the hot
core of the RG (or hot core of AGB star) or/and the accreting WD. This evolution finally could lead to the AIC of the WD (the PNe might be dissipated before the AIC), and this evolutionary pathway is also known as the symbiotic channel. In the following subsection, I will show the details for \textsc{MESA} simulation, and discuss the possible formation of NSs inside PNe from results of the WD-RG binary evolution.


\subsection{The WD - RG Binary stellar evolution with \textsc{MESA}}

The version 15140 of \textsc{MESA} code (i.e. \citealt{paxton2011, paxton2015, paxton2019}) is adopted to run the WD - RG star binary stellar evolution. For simulating the RG stars,
I set the hydrogen abundance, He abundance
and metallicity as $X = 0.70$, $Y = 0.28$ and $Z = 0.02$, respectively. For the other essential stellar evolution parameters,
I use $initial\_zfracs = 6$ and $kappa\_file\_prefix =$ `a09' to
call the opacity tables, and follow the Henyey theory of convection with $mixing\_length\_alpha=1.8$. 
With these parameters, I start the stellar evolution from a pre-main sequence model until the central helium fraction of the star is up to $\geq 0.98$, and it's radius is much larger than the original radius
when the star was at the MS stage. Then, I stop it and save the final model as the RG star model.

In order to show the possible formation of NSs inside PNe from the WD - RG binary evolution, with the \textsc{MESA} binary package I simulate the detailed evolution for a binary consisted of a non-magnetic WD with $1.2\,M_\odot$ and a RG star with $1.5\,M_\odot$, which has the 120 days of initial orbital period. The accretor ONeMg WD is treated as a point mass in the simulation, and I upload the saved RG star model which is built as described above. When the binary evolution starts to run, the angular momentum evolution is also taken into account (see \citealt{paxton2015}).
As discussed before, the mass transfer process has the key role in the binary evolution, especially for the possible formation of NSs inside PNe which is the main new point of this paper. If a star's radius is very closing to (or beyond) it's Roche lobe (RL) radius, the star starts to lose its mass. The star's effective RL radius is derived as shown in \cite{eg1983} ,
\begin{equation}
 R_{\rm RL} =(\frac{0.49q^{2/3}}{0.6q^{2/3} + {\rm ln}(1+q^{1/3})})a,
\end{equation}
where $a$ is the orbital separation and $q = M_{\rm RG}/M_{\rm WD}$. 
The Roche lobe overflow (RLOF) mass-transfer process is more complicated due to low surface gravities and extended atmospheres of giant stars. 
A more reasonable prescription for the mass transfer is studied by \cite{kr1990}. Thus, I adopt Kolb scheme given in the \textsc{MESA} code to compute the RLOF mass transfer $\dot{M_{\rm RL}}$ from the RG star,
\begin{multline}
{\dot{M}_{\rm RL}} =
-\dot{M}_{0}-2{\pi}F(q_2)\frac{R^3_{\rm RL}}{GM_{\rm RG}}\\
\times{\int^{P_{\rm RL}}_{P_{\rm ph}} {{\varGamma_1}^{1/2}{(\frac{2}{\varGamma_1 +1})}^{(\frac{\varGamma_1 +1}{2\varGamma_1 -2})}
{(\frac{\kappa_{\rm B} T}{m_{\rm p} \mu})}}\,\mathrm{d}P} ,
\end{multline}
where $\dot{M_{\rm RL}}$, $\varGamma_1$, $P_{\rm ph}$ and $P_{\rm RL}$ are the RLOF mass transfer rate, the first adiabatic exponent, the pressures at the photosphere and at the radius
when the radius of the donor is equal to its RL radius, respectively.
$T$, $\kappa_{\rm B}$ and $\mu_{\rm ph}$ are the temperature of the donor, Boltzmann constant and the mean molecular weight, respectively.
$\dot{M}_{0}$ is

\begin{equation}
\dot{M}_{0} = \frac{2\pi}{\rm exp(1/2)}F(q_2)\frac{R^3_{\rm RL, d}}{GM_d}{(\frac{\kappa_{\rm B} T_{\rm eff}}{m_{\rm p} \mu_{\rm ph}})^{3/2}}\rho_{\rm ph},
\end{equation}
where $T_{\rm eff}$ is the effective temperature of
the donor. $m_{\rm p}$, $\mu_{\rm ph}$ and $\rho_{\rm ph}$ are the proton mass, the mean molecular weight and density at its photosphere.
The fitting function $F({q_2})$ ($q_2 = M_{\rm accretor}/M_{\rm donor}$) is,
\begin{equation}
\begin{array}{ll}
F(q_2) =
1.23 + 0.5{{\rm log}(q_2)} & \textrm{$0.5\lesssim q_2 \lesssim 10$},
\end{array}
\end{equation}
Validation of $F({q_2})$ is same as in the MESA code. 
The following scheme is used for RG branch (RGB) stellar wind mass loss,
\begin{equation*}
\left.\begin{aligned}
\rm{cool\_wind\_RGB\_scheme = 'Reimers'}\\
\rm{cool\_wind\_AGB\_scheme = 'Blocker'}\\
\rm{RGB\_to\_AGB\_wind\_switch = 1d-4}\\
\rm{Blocker\_scaling\_factor = 0.0003d0}
\end{aligned}\right\}
\qquad \text{Stellar Wind}
\end{equation*}
I adopte typical ones described in the instrumental papers of MESA (e.g., \citealt{paxton2015}) for the other parameters or/and processes in the evolution.

One of decisive issues in this evolution is how much transferred mass actually accreted by the WD or how much of it is lost, and the mass accretion/loss is dominated by RLOF mass transfer rate. 
Whether the accreted hydrogen and helium could burn stably on the WD surface, it depends on the ${\dot{M}_{\rm RL}}$, and it alters the mass retention efficiency to ensure the mass growth of WD, so the mass growth rate of the WD can be calculated as,  
\begin{equation}
\dot{M}_{\rm{WD}} = \eta_{\rm H} \eta_{\rm He} {\dot{M}_{\rm RL}}.
\end{equation}
The prescriptions of \cite{hill2015} and \cite{kato2004} for the efficiency of
hydrogen burning ($\eta_{\rm H}$) and the mass accumulation efficiency of helium ($\eta_{\rm He}$) are adopted in the simulation.
The MESA simulation results of \cite{brook2017} show that the carbon burning has very little effect on the mass growth of WD.  The WD has almost no mass growth if nova bursts happen (i.e. \citealt{schaefer2025, schaefer2026}).


\begin{figure}
\centering
\includegraphics[totalheight=4.0in,width=3.5in]{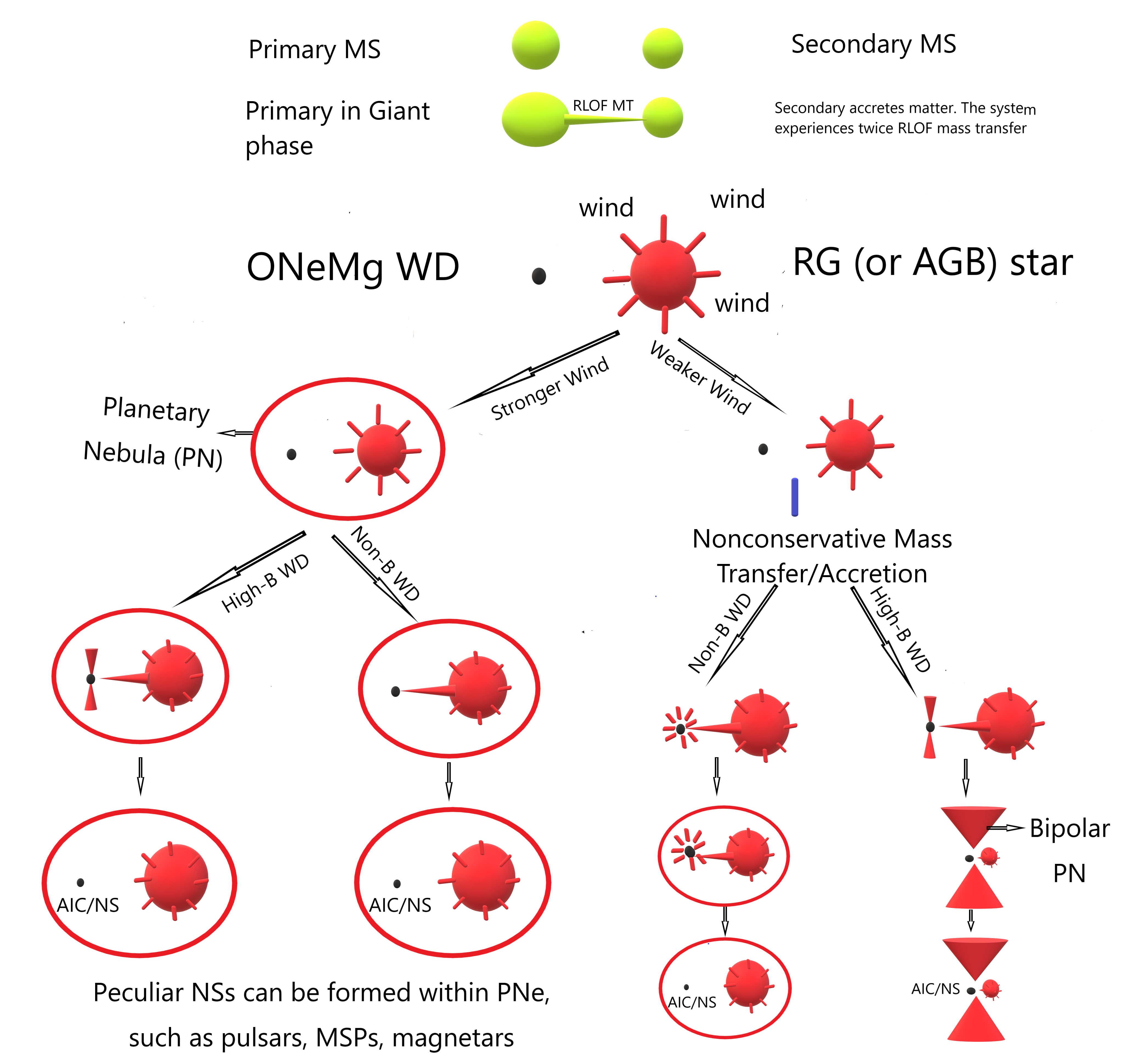}
\caption{Evolutionary pathways of WD binaries to form PNe (symbiotic nebulae) and the newborn neutron star. Abbreviations: MS: main sequence (star); RG: red giant; AGB: asymptotic giant branch; RLOF MT: Roche lobe overflow Mass transfer; WD: white dwarf; NS: neutron star.}
\end{figure}


\subsection{The \textsc{MESA} simulation result and discussion}

With the model introduced in the above subsection, I conducted the 1D  \textsc{MESA} simulation for the evolution of a WD-RG binary, where the ONeMg WD (with an initial mass of $1.2\,M_\odot$) can reach the Chandrasekhar limit mass, and the RLOF mass transfer from the RG star (with an initial mass of $1.5\,M_\odot$)  proceeds on a thermal timescale. As figure 2 shows,  the wind mass loss rate is less than
$10^{-12}\,\rm{M_\odot\,yr^{-1}}$ during the evolution, and the RLOF mass transfer rate dominates the mass accretion phase. Because the RLOF mass transfer rate is much higher, the hydrogen burning efficiency $\eta_{\rm H} $ (or mass retention efficiency $\eta_{\rm H} \eta_{\rm He}$) on the WD is between 0.4 and 1. Thus, the WD only accretes about the half of transferred mass, but the RG loses $\sim 0.4\,M_\odot$ in total when the WD reaches the limit mass (see Figure 2) and becomes a NS. The other part ($\sim 0.2\,M_\odot$) is ejected by WD, and the ejected mass probably escapes and surrounds the system, which may form a nebula (symbiotic nebula in this case) that could be heated by the very hot core of the RG and burning WD. 
Post-RG stars of mass $\leq 0.3\,M_\odot$ do not reach high enough effective temperatures to ionize the nebula even if are fully stripped of their envelope and evolve to become a WD (e.g. \citealt{hall2013}), and I find the core mass of giant companions in the AIC are massive to have high enough effective temperatures.
Therefore, the formation of a NS inside the symbiotic nebula is possible via the AIC from the WD binary evolution as demonstrated by the results of this simulation. 

If we compare the new result shown in this work with previous similar works, \cite{ablimit2023} have simulated the similar WD-RG binaries, but they did not even mention or discuss about the PNe. Let's take one WD-RG binary evolution from  \cite{ablimit2023} for the comparison, a binary given in Figure 3 of their paper, and parameters of this binary are as follows: the initial masses of the WD and RG donor are 1.2 $M_\odot$ and 1.3 $M_\odot$, and the initial period is 40 days. In the non-magnetic WD case, 
the RG donor transfers $\sim 0.3\,M_\odot$ in total to the WD via the RLOF, and $0.12\,M_\odot$ mass could escape
from the system ($\eta_{\rm H}<1.0$) while the WD accretes $0.18\,M_\odot$ and its mass reaches the Chandrasekhar limit mass and becomes a NS.
The large amount of escaped mass at the very late phase of the evolution may form a nebula around the binary before the AIC occurs, and the symbiotic nebula may be ionized by the hot core of the giant or/and the accreting WD.
If the WD in the same binary system is highly magnetized, the escaped mass at the late evolutionary phase (less than a time of $10^4$ year) could be even more due to the higher pole-mass transfer based on the \textsc{MESA} simulation results of  \cite{ablimit2023} (see also \cite{ablimitmaeda2019}).
It is possible that the large amount of transferred mass may be ejected from the two polar regions of the highly magnetized WD and may launch a pair of opposite jets, when finally it may form a bipolar symbiotic nebula.

\begin{figure}
\centering
\includegraphics[totalheight=2.5in,width=3.0in]{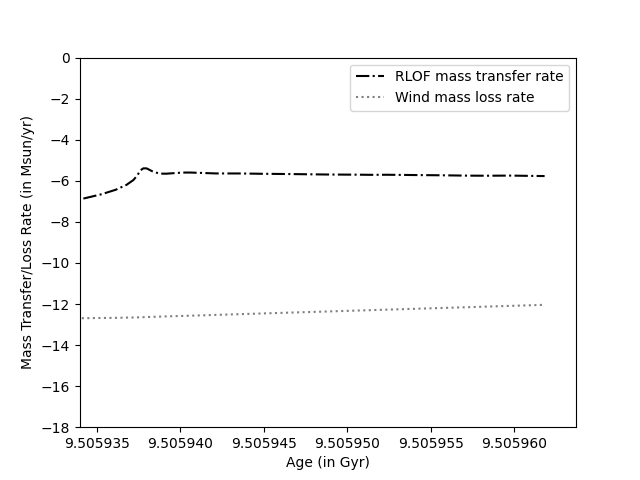}
\includegraphics[totalheight=2.5in,width=3.0in]{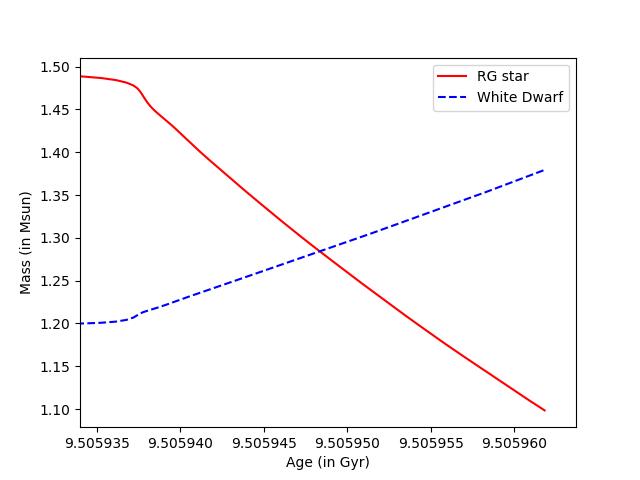}
\caption{Evolution of a WD - RG binary which may form PN (symbiotic nebula) and the newborn NS. Results are derived from a 1D \textsc{MESA} simulation of a binary consisting of a $1.2\,M_\odot$ WD and a $1.5\,M_\odot$ RG star in the 120 day orbit. Upper figure shows the RLOF mass transfer rate and wind mass loss rate of the RG star, and lower figure shows the mass evolution of the RG star and WD.}
\end{figure}

\begin{figure}
\centering
\includegraphics[totalheight=3.5in,width=3.5in]{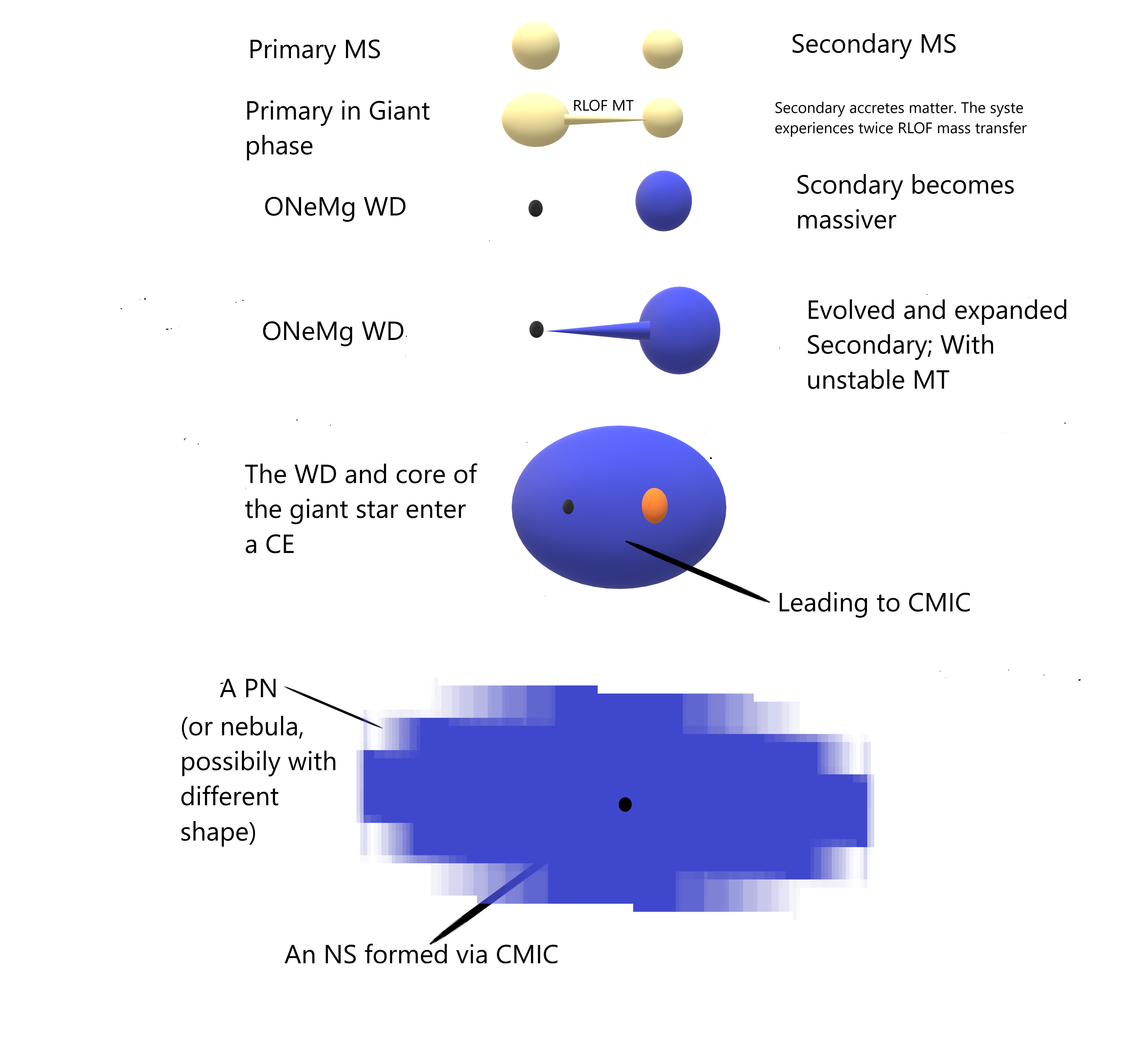}
\caption{A typical evolutionary pathway to form neutron stars inside PNe via the core-merger-induced collapse (CMIC). CE: common envelope; see the text and Figure 1 for other abbreviations.}
\end{figure}

\section{Formation of neutron stars with PNe via the CMIC}
\label{sec:concl}

\begin{figure}
\centering
\includegraphics[totalheight=3.5in,width=3.5in]{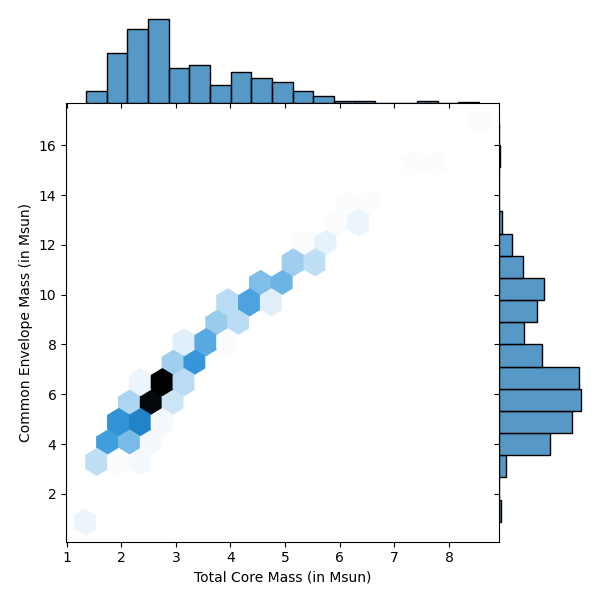}
\caption{A jointplot: distributions of masses of CE and total core from ONeMg WD  binaries during the CE phase. Total cores are the addition of the WDs and core of companion giant stars. WD binaries which lead to the merger of WD and RG's core inside the CE are selected with a condition of total core masses being larger than $1.44\,M_\odot$, and these systems are expected to form  neutron stars inside PNe via the CMIC.}
\end{figure}

The common envelop evolution (CEE) caused by the
unstable RLOF mass transfer has been proposed to form PNe with a rich variety of properties.
Merger of a CO WD companion with the core during the CEE might form a PN and might even trigger an type Ia supernova (SN) explosion during the PN phase
(e.g., \citealt{TsebrenkoSkoer2013, TsebrenkoSkoer2015, Cikotaetal2017, Chiotellisetal2020, Chiotellisetal2021}).
\cite{canals2018} has only mentioned one possibility of merger between a WD with the ONe core of an AGB during the CEE for the NS formation, while \cite{sabachsoker2014} mentioned only one possible way of NS formation through the merger from evolution of a WD and  a massive star (with the mass $>8.5\,M_\odot$) binary, and both of them did not mention PNe at all and did not study the mentioned two possibilities (it's not focus of their works) in details. I proposed and studied mergers between WDs and cores of all types of giants including subgiants, RG, AGB, stripped He star giants and massive giant stars as possible ways to form NSs in the CEE of WD binaries, and I referred it as the core-merger-induced collapse (CMIC) in \cite{Ablimitetal2022b} for the first time. Thus, the CMIC proposed by me includes more possible mergers which are studied in much more details, and it has more broader definition discussed with more different outcomes (i.e. millisecond pulsars, magnetars, Thorne-$\dot{\rm Z}$ytkow objects and black holes) comparing to all previous works. It is worth noting that the CMIC in \cite{Ablimitetal2022b} and all other previous works did not even mention/discuss the CEE of WD binaries with the formation of NSs and PNe together.  In this work, I first present and study the formation of NSs via the CMIC inside PNe after the CEE started from the evolution of ONeMg WD - RG or AGB star binaries as illustrated in Figure 3.
A massive and hot WD may be engulfed by the envelope of its giant companion due to the unstable RLOF mass transfer, and the system enters a CE phase. The WD may merge with the hot core of its giant companion inside the CE, and the CMIC may form a single peculiar NS inside a nebula. By adopting the binary population synthesis technique and its results, 
I will discuss more about it in the following subsection.


\subsection{The model set}

In order to explore the possibility of NS formation inside PNe via the CMIC senario,  $10^7$ primordial binaries consisting of two ZAMS stars are simulated with the updated \textsc{BSE} binary population synthesis (BPS) code (see \citealt{hurley2002, Ablimitetal2022b}). The fast BPS binary evolution has main processes like mass transfer, stellar winds, tidal interaction, angular momentum evolution, supernova explosions and natal kicks, common envelope (CE) evolution, magnetic braking, GW radiation, merger, etc.
I will briefly describe the main treatments in this model in the following.

For the initial mass distribution of primary stars that have higher masses than that of their companions in binary systems,  I adopt the initial mass function of \cite{kroupa1993},
\begin{equation}
f(M_1) = \left\{ \begin{array}{ll}
0 & \textrm{${M_1/M_\odot} < 0.1$}\\
0.29056{(M_1/M_\odot)}^{-1.3} & \textrm{$0.1\leq {M_1/M_\odot} < 0.5$}\\
0.1557{(M_1/M_\odot)}^{-2.2} & \textrm{$0.5\leq {M_1/M_\odot} < 1.0$}\\
0.1557{(M_1/M_\odot)}^{-\alpha} & \textrm{$1.0\leq {M_1/M_\odot} \leq 150$},
\end{array} \right.
\end{equation}
where $\alpha = 2.7$ in this work. The mass distribution of secondary stars is calculated by the distribution of the initial mass ratio,
\begin{equation}
n(q) = \left\{ \begin{array}{ll}
0 & \textrm{$q>1$}\\
\mu q^{\nu} & \textrm{$0\leq q < 1$},
\end{array} \right.
\end{equation}
where $q=M_2/M_1$, $\mu$ is the normalization factor for the assumed power law distribution with the index $\nu$. A flat distribution ($\nu = 0$ and $n(q)=$constant) for the initial mass ratio distribution is adopted. The following formalism is used for the distribution of the initial orbital separation, $a_{\rm i}$,
\begin{equation}
n(a_{\rm i}) = \left\{ \begin{array}{ll}
0 & \textrm{$a_{\rm i}/R_\odot < 3$ or $a_{\rm i}/R_\odot > 10^{6}$}\\
0.078636{(a_{\rm i}/R_\odot)}^{-1} & \textrm{$3\leq a_{\rm i}/R_\odot \leq 10^{6}$} \ .
\end{array} \right.
\end{equation}
The uniform (flat) initial eccentricity distribution is assumed in a range between 0 and 1. The metallicity is fixed as $Z=0.02$.

The CE evolution is an unsolved phase in the binary evolution process, and it is the key physical process in this work.
I adopt the widely used $\alpha$-formalism (\citealt{webbink1984}) for the CE phase,
\begin{equation}
E_{\rm{bind}} = {\alpha_{\rm CE}} \Delta E_{\rm orb} \ ,
\end{equation}
where $E_{\rm{bind}}$, $\alpha_{\rm CE}$, and $\Delta E_{\rm orb}$ are the binding energy of the envelope,
the efficiency parameter, and the change in
the orbital energy during the CE phase, respectively. For the binding energy of the envelope,
\begin{equation}
E_{\rm{bind}} = - \frac{GM_1 M_{\rm{en}}}{\lambda {R}_1},
\end{equation}
where $M_1$, $M_{\rm{en}}$ and ${R}_1$ are the total mass, envelope
mass, and radius of the primary star, respectively.  The two stars may merge during the CE phase or survive from the CE; for details of the criterion for surviving or merging during the CE phase, see \cite{hurley2002}. In this CE model, the efficiency parameter ($\alpha_{\rm CE}$) and the binding energy parameter ($\lambda$) are very important for the outcome of binary evolution. In this work, I set $\alpha_{\rm CE} = 1.0$ and $\lambda = 0.5$ in a widely considered basic way.

The rapid remnant-mass model of \cite{fryer2012} in the subroutine
hrdiag of the updated BSE code is used to calculate the SN remnant. I correct a typo in \cite{fryer2012} in the code [corrected as $a1 = 0.25 - 1.275/(M_1 - M_{\rm proto})$), and the proto-compact object mass is
set as $M_{\rm proto} = 1.0\,M_\odot$]. For the natal SN kick velocity, imparted to the newborn NS at its birth, I adopt Maxwellian distributions
with a velocity dispersion of $\sigma_{\rm k} = 265 \,\rm km\,s^{-1}$ (\citealt{hobbs2005}) for Core-collapse SNe  and $\sigma_{\rm k} = 40 \,\rm km\,s^{-1}$ for electron-capture SNe or AIC. I adopt default treatments for other processes and parameters such as the wind mass loss, critical mass ratio and tidal effect (see \cite{hurley2002} for more details).


\subsection{The BPS result and discussion}

Here, I examine a scenario in which, following the formation of a CE system containing an ONeMg WD, the WD and the donor star's core spiral inward. If the CE cannot be successfully ejected,  the likely outcome is the collapse of the WD into a more compact object such as a NS. I explore the evolutionary pathway involving a CE phase, under the assumption that the CMIC occurs within the CE if the combined mass of the ONeMg WD and the giant star's core exceeds the Chandrasekhar limit (1.44 $M_\odot$; the hot cores of giant campanions are massive than $0.3\,M_\odot$). Thus, the combined core mass and CE mass are key parameters to show the possible NS formation inside a nebula.

Based on the BPS results,  the Galactic formation rate of CMIC inside PNe is $\sim 3.6\times10^{-4} \,{\rm yr^{-1}}$ if a star formation rate of 2 $M_\odot \,{\rm yr^{-1}}$ is adopted. The CE mass and the total core mass of ONeMg WDs and their giant companions in binaries that lead to CMIC are shown in Figure 4.  The figure shows that two mass distributions have wide enough ranges for forming both nebulae and NSs. The hot WDs or/and hot giant cores might reach the effective temperatures of 30000 K that are necessary to ionize the inner part of or whole CE during the spiraling-in process, and the collapse explosion might cause the expansion of a partially or fully ionized CE to form a nebula. In another way, the new born NS which is expected to be hot (and also magnetically active) may power the expanding nebula. The results derived from the BPS simulations of this work and \cite{Ablimitetal2022b} show the importance of this type of peculiar object. Observations on pulsar wind nebulae inside the expanding PNe may be helpful to understand this kind of evolution.


\section{Conclusion}
\label{sec:concl}

The formation of NSs has been extensively studied in the previous works; however, the possibility of NS formation within a planetary nebula has not been previously addressed. In this paper, I propose two potential mechanisms for NS formation inside planetary nebulae: (1) the AIC of ONeMg WDs occurring within planetary nebulae--referred to here as symbiotic nebulae, and (2) the CMIC, which may take place during the CEE of ONeMg WD binaries, leading to the formation of both an NS and a planetary nebula. With the advent of next-generation ground- and space-based telescopes, compelling evidence for these scenarios may soon emerge, offering deeper insights into binary stellar evolution and rare phenomena discussed in this work such as WD collapses and explosions inside the planetary nebulae.

\section*{Acknowledgments}

I would like to thank Noam Soker for his useful comments that improved the text.

\section*{Data Availability}

The data underlying this article, and the necessary MESA/BPS code files for reproducing our models will be shared on reasonable request to the corresponding author.





 






\bsp	
\label{lastpage}
\end{document}